\documentclass[preprint,showpacs,amsmath]{revtex4}

\usepackage{graphicx}
\usepackage{dcolumn}
\usepackage{bm}
\begin{document}

\def\s#1{\slash\!\!\!{#1}}
\def\bee{\begin{eqnarray}}
\def\eee{\end{eqnarray}}
\def\nn{\nonumber\\}
\def\n{\nonumber}
\def\fr#1#2{\frac{#1}{#2}}
\def\Tr{\makebox{Tr}}
\title{$\eta-\eta^\prime$ Mixing and NLO Power Correction}
\author{Tsung-Wen Yeh}
\email{twyeh@cc.nctu.edu.tw}
\affiliation{%
Institute of Physics, National Chiao-Tung University,Hsinchu 300, Taiwan
}%
\begin{abstract}
The next-to-leading-order (NLO) $O(1/Q^4)$ power correction for $\eta\gamma$ and $\eta^\prime\gamma$ form factors are evaluated and employed to explore the $\eta-\eta^\prime$ mixing. The parameters of the two mixing angle scheme are extracted from the data for form factors, two photon decay widths and radiative $J/\psi$ decays. The $\chi^2$ analysis gives the result: $f_{\eta_1}=(1.16\pm0.06)f_\pi, f_{\eta_8}=(1.33\pm0.23)f_\pi, \theta_1=-9^\circ\pm 3^\circ, \theta_8=-21.3^\circ\pm 2.3^\circ$, where $f_{\eta_{1(8)}}$ and $\theta_{1(8)}$ are the decay constants and the mixing angles for the singlet (octet) state. In addition, we arrive at a stringent range for $f_{\eta^\prime}^c:-10$ MeV$\le f_{\eta^\prime}^c\le -4$ MeV.
\end{abstract}

\pacs{12.38.Bx,14.40.-n}

\maketitle

\section{INTRODUCTION}
Recently, the next-to-leading order (NLO) power correction has been shown an important role in understanding the exclusive processes $\gamma^*\pi\to\gamma$ \cite{twy1} and $\gamma^*\pi\to\pi$ \cite{twy2}. 
The method for calculating the NLO power correction is called the collinear expansion \cite{twy1,twy3,Ellis:1982wd,Ellis:1983cd,Qiu:1990dn}. This power expansion method is compatible with PQCD factorization \cite{twy1,twy2}, which demonstrates the amplitude as convolutions of perturbatively calculable hard scattering amplitudes (the hard function) and nonperturbative hadronic wave functions (the soft function). Furthermore, it is a Feynman diagram approach such that the partonic interpretation for the NLO power correction can be preserved. 

The asymptotic limit of the $\pi\gamma$ transition form factor to be $2f_\pi/Q^2$ with $f_\pi=93$ MeV, is about $15\%$ higher than the upper end of the CLEO data \cite{Gronberg:1998fj}. This deviation can be successfully explained by NLO power correction. In addition, NLO power correction can well describe the low energy portion of the CLEO data. 
Our purpose in this paper is to generalize the approach for the $\pi\gamma$ form factor to the $\eta\gamma$ and $\eta^\prime\gamma$ form factors. The first problem we shall face is that there are many independent degrees of freedom associated with the $\eta$ and $\eta^\prime$ lowest valence Fock states.  The $\eta$ and $\eta^\prime$ mesons are admixtures of the $SU(3)_F$ octet and singlet states, this counts eight quantities: four wave functions and four related decay constants. The $\eta$ and $\eta^\prime$ mesons can also receive contributions from the $U(1)_A$ anomaly to have intrinsic heavy quark and gluon contents. To reduce the number of independent degrees of freedom, we invoke phenomenological constraints and physical assumptions over the wave functions and decay constants. 

For the phenomenological constraints, we shall employ the large momentum transfer data for $\eta\gamma$ and $\eta^{\prime}\gamma$ transition form factor \cite{Gronberg:1998fj,Acciarri:1998yx,Behrend:1991sr,Aihara:1990nd,Berger:1984xk}, the two-photon decay widths of the $\eta$ and $\eta^\prime$ mesons and the ratio $R_{J/\psi}$ of the $J/\psi\to\eta\gamma$ and $J/\psi\to\eta^\prime\gamma$ decay widths. As for the physical assumptions we shall invoke the $SU(3)_F$ octet-singlet mixing scheme. In this mixing scheme, both $\eta$ and $\eta^\prime$ are linear combinations of $\eta_8$ and $\eta_1$, the octet and singlet states in the $SU(3)_F$ representation. The mixing is controlled by the mixing angle. In the one mixing angle scheme, in which the octet and singlet meson states have different decay constants $f_{\eta_8}$ and $f_{\eta_1}$ and share a common mixing angle $\theta_8=\theta_1=\theta$, the mixing angle $\theta$ is in range of $-20^{\circ}\le \theta\le -10^{\circ}$ \cite{Gilman:1987ax}. In recent years, many evidences \cite{Leutwyler:1998yr,Feldmann:1998vc,Feldmann:1998yc} have indicated that a two mixing angle scheme, in which the octet and singlet mixing angles $\theta_8$ and $\theta_{1}$ can take different values, is more general than the one mixing angle scheme. The $\eta\gamma$ and $\eta^\prime\gamma$ form factors have been investigated by either one mixing angle scheme \cite{Jakob:1996,Anisovich:1996hh,Ametller:1992} or two mixing angle scheme\cite{Feldmann:1998vc}. We would like to make a more refined analysis for these form factors by including NLO power correction. 

 In this paper, we shall investigate the $\eta\gamma$ and $\eta^\prime\gamma$ form factors and the mixing pattern of $\eta-\eta^\prime$ system by employing PQCD formula with NLO power correction in the standard hard scattering approach. We shall employ the collinear expansion to derive NLO ($O(Q^{-4})$) power correction for $\eta\gamma$ and $\eta^\prime\gamma$ form factors in Sec.~II.  In Sec.~III we shall analyze the high momentum transfer data for form factors, the two photon decay widths and the ratio $R_{J/\psi}$. The values of the mixing parameters are determined from a $\chi^2$ analysis for the data. Sec.~IV is devoted to conclusions.

\section{$\eta\gamma$ and $\eta^\prime\gamma$ form factors and collinear expansion}
Our strategy in calculating the power corrections to the $\eta$ and $\eta^\prime$ meson-photon transition form factor is to invoke the collinear expansion \cite{twy1,Ellis:1982wd,Ellis:1983cd,Qiu:1990dn}. 
For simplicity, we shall first ignore the meson mass effects. That is we choose the momentum of the initial state meson, $P_1$, and that of the final state photon, $P_2$,  as
\bee
P_1^\mu&=&(Q,\frac{M_P^2}{2Q},0_\perp)\equiv p^\mu+\frac{M_P^2}{2}n^\mu\approx p^\mu\ ,\nn
P_2^\mu&=&(0,\frac{Q}{2},0_\perp)\equiv \frac{Q^2}{2} n^\mu\ ,
\eee
such that the virtual photon has momentum $q=P_2-P_1$ with virtuality $q^2=-Q^2$ to make PQCD applicable. Vectors $p$ and $n$ are in the $+$ and $-$ directions in the light-cone reference frame and have properties $p^2=n^2=0$ and $p\cdot n=1$. $M_P$ denotes the mass of the initial state meson.
For the Feynman diagrams displayed in Figs.~1(a) and (b), the amplitudes are written as 
\bee
A(P_1,P_2)=\int \frac{d^4 k}{(2\pi)^4}\Tr[H(k,P_2,Q^2)\Phi(k,P_1,Q^2)]
\eee
where the trace is taken over the color and spin indices and the meson DA $\Phi(k,P_1,Q^2)$ has expression for $P(=\eta, \eta^\prime)$ meson 
\bee 
\Phi(k,P_1,Q^2)=\int\frac{d^4 k}{(2\pi)^4}\int d^4 z e^{ik\cdot z}\langle 0|\bar{q}(0)q(z)|P\rangle\ .
\eee
We assign the loop momentum $k$ for the valence antiquark and let it flow into the hard function. The hard function $H(k,P_2,Q^2)$ contains two parton photon interaction vertices and one virtual internal parton propagator. The amplitude $A$ contains leading, next-to-leading and higher twist contributions. The quantity twist is understood as an effective twist for nonlocal operators and is not exactly the same with the usual twist defined for local operators. The twist has different meanings for the hard and soft functions. For the hard function, the twist is defined as the power of the inverse of the photon virtuality $Q$, and, for the soft function, the twist represents the power of the small scale $\Lambda$ with magnitude of order $\Lambda_{QCD}$.  By employing collinear expansion, we can systematically separate the leading twist (LT) contributions from the next-to-leading twist (NLT) contributions. The LT contributions are from collinear loop momentum $\hat{k}=xp$. It is therefore convenient to parameterize the loop momentum $k$ into
\bee
k^{\mu}=x p^\mu + \frac{k^2+k_\perp^2}{2x }n^\mu + k_\perp^\mu\ ,
\eee
where $k$ contains on-shell part
\bee
k^\mu_L=x p^\mu + \frac{k_\perp^2}{2x }n^\mu + k_\perp^\mu\
\eee 
and off-shell part 
\bee
k^\mu_S=\frac{k^2}{2x }n^\mu\ .
\eee
In the first step, we expand the hard function $H(k)$ with respect to $\hat{k}$ as
\bee\label{hexp}
H(k)=H(k=\hat{k})+\frac{\partial H(k)}{\partial k^\alpha}\Bigg|_{k=\hat{k}}(k-\hat{k})^\alpha+\cdots
\eee

With the help of $k_L$ and $k_S$, we can factorize the loop parton propagator $F(k)=-i/(\s{k}-i\epsilon)$ into its long distance part, $F_L(k)$, and short distance part (the special propagator defined in \cite{Qiu:1990dn}), $F_S(k)$, which take expressions 
\bee
F_L(k)&=&\frac{-i\s{k}_L}{k^2-i\epsilon}\;\; ,\;\; 
F_S(k)=\frac{-i\s{n}}{2k\cdot n-i\epsilon}\ .
\eee
The propagators $F_L(k)$ and $F_S(k)$ have different physical meanings. To see this, it is amount to consider their propagations on the light cone. The integrals of $F_L(k)$ and $F_S(k)$ over $k^+=k\cdot n$ give 
\bee
f_L(\eta,\lambda)&=&\int\frac{dk\cdot n}{(2\pi)}e^{ik\cdot n(\eta-\lambda)}F_L(k)\propto\theta(\eta-\lambda)\ ,\nn
f_S(\eta,\lambda)&=&\int\frac{dk\cdot n}{(2\pi)}e^{ik\cdot n(\eta-\lambda)}F_S(k)\propto\delta(\eta-\lambda)\ ,
\eee
where $\eta$ and $\lambda$ mean the light-cone distances in the $-$ direction.
It is obvious that $F_S(k)$ is not propagating on the light cone.
This means that $F_S(k)$ should be included into the hard function. By dimensional counting, $F_S(k)$ is of order $O(Q^{-1})$. Therefore, including one $F_S(k)$ into the hard function then increases one twist order for the hard function. 

There are different effects as $F_L(k)$ and $F_S(k)$ act on the spin structures of hard function, the terms proportional to $\s{p}$ or $\s{n}$. As $F_L(k)$ acts on $\s{p}$, its collinear part vanishes and non-collinear parts are retained  
\bee\label{cl1}
F_L(k)\s{p}=-F(k)(k-\hat{k})^\alpha(i\gamma_\alpha)F_S(k)\s{p}
\eee
where minus sign comes from the anti-particle propagator.
The vertex $i\gamma_\alpha$ and short distance propagator $F_S(k)$ are then absorbed into the hard function. The factor $(k-\hat{k})^\alpha$ is included into the soft function to become a coordinate derivative on the quark fields. As $F_L(k)$ acts on $\s{n}$, its collinear part contributes to leading order. The short distance propagator $F_S(k)$ only serves to introduce the interaction term $\bar{q}\s{A}q$ for $\s{p}$ vertex, where $A^\alpha$ denote the gluon fields. The total effects of $F_L(k)$ and $F_S(k)$ acting on $\s{p}$ are to include one $i\gamma_\alpha$ and one $F_S(k)$ into the hard function and to absorb the factor $(k-\hat{k})^\alpha$ and gauge fields $A^{\alpha}$ into the soft function to become a covariant derivative, $D^{\alpha}=i\partial^{\alpha}-g A^{\alpha}$ with $g$ the strong coupling.

The contributions from the second term of Eq.~(\ref{hexp}) and from Figs.~1.(c) and (d) are of twist-6 or higher twist and will not be considered in below discussions. The reason is that the possible non-vanishing components of $\gamma_\alpha$ in ${\partial H(k)}/{\partial k^\alpha}$ are  $\alpha=+$ or $-$, but both vanish as ${\partial H(k)}/{\partial k^\alpha}$ contract with $(k-\hat{k})^\alpha$ or $\langle 0|\bar{q}A^{\alpha}q|P\rangle$. We substitute the first term of Eq.~(\ref{hexp}) into the integral with the soft function and apply the identity
\bee
\int dx \delta(x-k\cdot n)=1
\eee
to convert the loop momentum integral into the fraction variable integral. The amplitude then becomes, approximately,
\bee
A(P_1,P_2)\approx\int dx \Tr[H(x)\Phi(x)]
\eee
which contains LT and NLT contributions. The meson DA $\Phi(x)$ has the expression
\bee
\Phi(x)=\int_0^\infty\frac{d\lambda}{2\pi}e^{ix\lambda}\langle 0|\bar{q}(0)q(\lambda n)|P\rangle\ .
\eee
We now discuss how to separate the LT from the NLT contributions for amplitude $A$. Due to the fact that the final state photon is real and has transverse polarization, the hard function $H(x)$ can have spin structures: $\gamma_\perp\s{n}\gamma_\perp$,  $\gamma_\perp\s{p}\s{n}$ $\gamma_\perp\s{n}\s{p}$ and $\gamma_\perp\s{p}\gamma_\perp$ where $\gamma_\perp=\gamma^\alpha$ with $\alpha=1,2$. The first spin structure leads to LT contribution, while the second and third ones result in the NLT contributions. The last spin structure would lead to next-next-to-leading twist contribution and will not be considered below. To calculate the NLT contributions, we need to apply Eq.~(\ref{cl1}) to extract the contributions from non-collinear loop momentum. As a result, we get the amplitude up to NLT as 
\bee\label{nloa}
A\approx \int dx \Tr[H(x)\Phi(x)]+ \int dx  \Tr[H_{\alpha}(x)w_{\alpha^\prime}^{\alpha} \Phi^{\alpha^\prime}(x)]
\eee
where the first term of the right hand side of Eq.~(\ref{nloa}) comes from the Feynman diagrams shown in Fig.~1(a) and (b) and the second term from those diagrams shown in Fig.~2. The tensor $w_{\alpha^\prime}^{\alpha}$ is defined as $w_{\alpha^\prime}^{\alpha}=g_{\alpha^\prime}^{\alpha}-p^{\alpha}n_{\alpha^\prime}$. The NLT hard function $H_{\alpha}(x)$ is defined as
\bee
H_{\alpha}(x)&=&(i\gamma_\alpha)\frac{-i\s{n}}{2x}H(x)+H(x)(i\gamma_\alpha)\frac{i\s{n}}{2(1-x)}\ 
\eee
and the NLT meson DA $\Phi^\alpha(x)$ has expression
\bee
\Phi^\alpha(x)=\int_0^1 dx_1\int_0^\infty\frac{d\lambda}{2\pi}\int_0^\infty\frac{d\eta}{2\pi}e^{i(x_1-x)\eta}e^{ix\lambda}\langle 0|\bar{q}(0)D^\alpha(\eta n)q(\lambda n)|P\rangle\ .
\eee
The factorization of momentum integral is finished. To complete the factorization, we still need to perform the factorizations of the color and spin indices. To separate the color indices, we take the convention that the color factors of the hard function are extracted and absorbed into the soft function. As for the spin indices, we employ the expansion of the soft functions into their spin components
\bee
\Phi&=&\sum_{\Gamma}\Gamma\phi^{\Gamma}\ ,\nn
\Phi^{\alpha}&=&\sum_{\Gamma}\Gamma\phi^{\Gamma;\alpha}\ , 
\eee
where $\Gamma$ denotes gamma matrix and $\phi^{\Gamma(;\alpha)}$ is the related spin component of the distribution amplitude. For a given order of $1/Q^2$, we choose the component $\phi^{\Gamma(;\alpha)}$ with lowest twist. The determination of the lowest twist $\phi^{\Gamma(;\alpha)}$ can be done as follows. Firstly, we notice that the tensor structure of $\phi^{\Gamma(;\alpha)}$ can be expressed in terms of $p$, $n$, $d^{\alpha\beta}_\perp=g^{\alpha\beta}-p^{\alpha}n^{\beta}$ and $ \epsilon^{\alpha\beta}_{\perp}=\epsilon^{\alpha\beta\gamma\lambda}p_{\gamma}n_{\lambda}$. The vectors $p$ and $n$ have dimensions $[p]=1$ and $[n]=-1$ with respect to the hard scale $Q$. Secondly, note that the matrix element for the soft function is written as 
\bee
\Phi\sim \langle 0|\overline{q}q|P\rangle\ , \nn
\Phi^{\alpha}\sim \langle 0|\overline{q}D^{\alpha} q|P\rangle\ .
\eee
By the above facts, we can derive a power counting rule as follows. Consider the $\phi^{\mu_1\cdots\mu_F;\alpha_1\cdots\alpha_B}$ has the fermion index $F$ and the boson index $B$. The fermion index $F$ arise from the spin index factorization for $2F$ fermion lines connecting the soft function and the hard function and the boson index $B$ denotes the $n_D$ power of  momenta in previous collinear expansion and the $n_G$ gluon lines as $B=n_D+n_G$. We may write
\bee\label{pc1}
\phi^{\mu_1\cdots\mu_F;\alpha_1\cdots\alpha_B}=\sum_i \Lambda^{\tau_i-1}e^{\mu_1\cdots\mu_F;\alpha_1\cdots\alpha_B}_i\phi^i
\eee
where $\Lambda$ denotes a small scale associated with DA. Spin polarizers $e_i$ denote the combination of  vectors $p^\mu$, $n^\mu$ and $\gamma^\mu_\perp$. Variable $\tau_i$ represents the twist of DA $\phi^i$. The restrictions over polarizers $e^{\mu_1\cdots\mu_F;\alpha_1\cdots\alpha_B}_i$ are 
\bee\label{pc2}
n_{\alpha_j}e^{\mu_1\cdots\mu_F;\alpha_1\cdots\alpha_j\cdots\alpha_B}_i=0
\eee
which are due to the fact that polarizers $e_i$ are always projected by  $w^{\alpha}_{\alpha^\prime}$.
The dimension of $\phi^{\mu_1\cdots\mu_F;\alpha_1\cdots\alpha_B}$ is determined by dimensional analysis
\bee\label{pc3}
d(\phi)=3F+B-1
\eee 
By equating the dimensions of both sides of Eq.(\ref{pc1}), one can derive the minimum of $\tau_i$ 
\bee\label{pc4}
\tau_i^{\min} =2F+B+\frac{1}{2}[1-(-1)^B]\ .
\eee
It is obvious from Eq.(\ref{pc4}) that there are only finite numbers of fermion lines, gluon lines and derivatives contributes to a given power of $1/Q^2$.

We now demonstrate that the collinear expansion is compatible with the conventional approach for proving the PQCD factorization at one loop order of radiative correction. To show this, we consider the radiative correction for $H(x)$ as displayed in Fig.~3(a). If the radiative gluon in Fig.~3(a) is collinear with momentum $l^{\alpha}=(l^+,l^-,l_\perp^i)=(Q,\lambda^2/Q,\lambda)$ , where $\lambda\ll Q$, the lower virtual antiquark has the momentum $(k-l)\sim(\xi Q,\lambda^2/Q,\lambda)$ with $k=xP_1$. It is obvious that the virtual antiquark in the collinear region behaves similarly to the loop antiquark in the tree amplitude. The collinear expansion for Fig.~3(a) in the collinear region is the same as the expansion for the tree diagram Fig.~1(a) in the leading configuration. To demonstrate this, we write the integrand for Fig.~3(a) as
\bee
I_{3a}&=&\int\frac{d^4l}{(2\pi)^4}\frac{g^2e^2\Gamma^{\beta\lambda}(l)}{l^2}\Tr[\gamma_{\beta}F(l_1)\s{\epsilon}F(l_3)\gamma^\mu F(l_2)\gamma_\lambda]\nn
&\equiv&\int\frac{d^4l}{(2\pi)^4} \Tr[H_{3a}(l)\Phi_{3a}(l)]
\eee
where we have defined $l_1=k-l, l_2=(P_1-k+l)$, $l_3=P_2-k+l$,
\bee
H_{3a}(l)&=&-ie^2[\s{\epsilon}F(l_3)\gamma^\mu]\ ,\nn
\Phi_{3a}(l)&=&\frac{ig^2\Gamma^{\beta\lambda}(l)}{l^2}[\gamma_{\beta}F(l_1) F(l_2)\gamma_\lambda]\ ,
\eee
and $F(l_i)=1/\s{l}_i$. We, firstly, expand $H_{3a}(l)$
\bee
H_{3a}(l)=H_{3a}(l=\hat{l})+\frac{\partial H_{3a}(l)}{\partial l^\lambda}|_{l=\hat{l}}(l-\hat l)^\lambda
\eee
with $\hat{l}=(x-\xi) P_1$. Repeating the same considerations for the expansion of the tree amplitude, we can recast $I_{3a}$ into
\bee
I_{3a}=\int d\xi [H^{(0)}_0(\xi)\Phi_{3a}(\xi)]+\int d\xi [H^{(0)}_\alpha(\xi) w_{\alpha^\prime}^\alpha\Phi_{3a}^{\alpha^\prime}(\xi)]+\cdots\ ,
\eee
where
\bee
\Phi_{3a}(\xi)&=&\int\frac{d\lambda}{(2\pi)}\int\frac{d^4 l}{(2\pi)^4}e^{i\lambda(\xi+x- l\cdot n)}[\frac{ig^2\Gamma^{\beta\lambda}(l)}{l^2}\gamma_{\beta}F(l_1) F(l_2)\gamma_\lambda]\ ,\nn
\Phi_{3a}^{\alpha^\prime}(\xi)&=&\int\frac{d\lambda}{(2\pi)}\int\frac{d^4l}{(2\pi)^4}e^{i\lambda(\xi+x-l\cdot n)} [\frac{ig^2\Gamma^{\beta\lambda}(l)}{l^2}\gamma_{\beta}F(l_1) l^{\alpha^\prime}F(l_2)\gamma_\lambda]\ .
\eee
It is obvious that both $\Phi_{3a}(\xi)$ and $\Phi_{3a}^{\alpha^\prime}(\xi)$ are collinear divergent for collinear $l$. We introduce corresponding soft functions $\Phi_0^{(1)}$ and $\Phi_1^{(1)}$ to absorbe $\Phi_{3a}(\xi)$ and $\Phi_{3a}^{\alpha^\prime}(\xi)$. The corresponding tree level hard functions are denoted as $H_0^{(0)}$  and $H_1^{(0)}$, respectively. If the radiative gluons in Fig.~3(a) are soft, i.e. the gluons have momentum $l=(l^+,l^-,l_\perp)\sim(\lambda,\lambda,\lambda)$, there are no effects on power expansion. This is because the eikonal approximation up to $O(1/Q^{6})$ can be applied to factorize the soft gluons from the valence quark propagator.  The other two particle reducible diagrams Figs.~3(b) and (c) can be dealt with similarly. It is noted that the double logarithms in Figs.~3(a) to (c) arising from the mixing contributions from the soft and collinear divergences are cancelled each other. In light-cone gauge $n\cdot A=0$, Figs.~3(d) and (e) in collinear region are more suppressed than Figs.~3(a) to (c) in collinear region by, at least, $1/Q^{2}$. After subtracting the soft contributions (the soft and collinear divergences) from the one loop radiative correction diagrams Figs.~3(a) to (c), we can obtain the one loop corrected hard functions (LO and NLO) $H_0^{(1)}$  and $H_1^{(1)}$, separately. The analysis for the radiative corrections to $H_{\alpha}(x)$ is simple, since it only involves radiative correction and has no need to consider collinear expansion. The diagrams for the radiative corrections to $H_{\alpha}(x)$ are shown in Fig.~4. As a result, up to first order in radiative and power corrections, we can arrive at the factorized amplitudes as 
\bee
A^{(0)}+A^{(1)}\approx (H^{(0)}_0+H_0^{(1)})\otimes (\Phi_0^{(0)}+\Phi_0^{(1)})+(H_1^{(0)}+H_1^{(1)})\otimes(\Phi_1^{(0)}+\Phi_1^{(1)})
\eee
where the superscript indices $i (i=0,1)$ denote the order of radiative correction and the subscript indices $j (j=0,1)$ mean the order of power correction. The notation $\otimes$ represents the convolution integral and the trace over the color and spin indices.
To prove PQCD factorization, we need to generalize the one loop factorization to arbitrary orders. It can be done straightforwardly \cite{twy4}.

For convenience, we may write the amplitude as
\begin{eqnarray}\label{pi3}
A(\gamma^* P\to\gamma)=-ie^2\epsilon_{\mu\alpha\beta\lambda}P_1^{\alpha}P_2^{\beta}\epsilon^{\lambda}F_{P\gamma}(Q^2)\ ,
\end{eqnarray}
where $\epsilon^{\lambda}$ denotes the polarization vector of the final state photon. The form factors are expressed in terms of the octet and singlet components 
\bee\label{nloff}
F_{P\gamma}(Q^2)=\sum_{i=8,1}a^{P_i}_PF_{P_i\gamma}(Q^2)
\eee
where the expansion coefficients $a_P^{P_i}, P=\eta,\eta^\prime, i=8,1$ depend on the mixing scheme (see next section). Due to the $\eta-\eta^\prime$ mixing, we take the octet and singlet states as the basis for our investigation of the form factors. The superscript $i=8$ or $1$ denote the contributions from octet or singlet current (see below notation). 
The leading order of $F_{P_i\gamma}(Q^2)$ is calculated from Figs.1(a) and (b) and takes expression
\bee
F_{P_i\gamma}^{\makebox{\tiny LO}}(Q^2)=4C_i\int_0^1 dx \frac{\phi_{P_i}(x)}{Q^2x(1-x)}\ ,
\eee
where the charge factors are defined as $C_8=(e_u^2+e_d^2-2e^2_s)/\sqrt{6}$ and $C_1=(e_u^2+e_d^2+e^2_s)/\sqrt{3}$. The NLO of $F_{P_i\gamma}(Q^2)$ is evaluated from Fig.2
\begin{eqnarray}
F_{P_i\gamma}^{\makebox{\tiny NLO}}(Q^2)=-16C_i\int_0^1 dx \frac{[G_{P_i}(x)+\tilde{G}_{P_i}(x)(1-2x)]}{Q^4x(1-x)}\ .
\end{eqnarray}
We have taken the symmetry between the exchange of $x\leftrightarrow (1-x)$ for $\phi_{P_i}(x)$, $G_{P_i}(x)$ and $\tilde{G}_{P_i}(x)$. 
The relevant DAs are expressed explicitly as follows
\begin{eqnarray}
\phi_{P_i}(x)&=&-i\frac{1}{4}\int_0^\infty \frac{d\lambda}{(2\pi)}e^{i\lambda x}
\langle 0|J^{i}_5(0,\lambda n)|P_i(P_1)\rangle\\
G_{P_i}(x)&=&-\frac{1}{8}\epsilon_\perp^{\alpha\beta}\int_0^1 d x_1\int_0^\infty \frac{d\lambda}{(2\pi)}\frac{d\eta}{(2\pi)}e^{i\eta( x_1-x)}
e^{i\lambda x}
\langle 0|J^{i}_{\alpha\beta}(0,\eta n,\lambda n)|P_i(P_1)\rangle\ ,\\
\tilde{G}_P^i(x)&=&-\frac{i}{8}d_\perp^{\alpha\beta}\int_0^1 d x_1\int_0^\infty \frac{d\lambda}{(2\pi)}\frac{d\eta}{(2\pi)}e^{i\eta( x_1-x)}
e^{i\lambda x}
\langle 0|J^{i}_{\alpha\beta 5}(0,\eta n,\lambda n)|P_i(P_1)\rangle\ .
\end{eqnarray}
where nonlocal currents are defined as
\bee
J^8_5(0,\lambda n)&=&\frac{1}{\sqrt{6}}[\bar{u}(0)\gamma_5\s{n}u(\lambda n)+\bar{d}(0)\gamma_5\s{n}d(\lambda n)-2\bar{s}(0)\gamma_5\s{n}s(\lambda n)]\ ,\nn
J^1_5(0,\lambda n)&=&\frac{1}{\sqrt{3}}[\bar{u}(0)\gamma_5\s{n}u(\lambda n)+\bar{d}(0)\gamma_5\s{n}d(\lambda n)+\bar{s}(0)\gamma_5\s{n}s(\lambda n)]\ ,\nn
J^8_{\alpha\beta}(0,\eta n,\lambda n)&=&\frac{1}{\sqrt{6}}[\bar{u}(0)\gamma_{\alpha}D_{\beta}(\eta n)u(\lambda n)+\bar{d}(0)\gamma_{\alpha}D_{\beta}(\eta n)d(\lambda n)-2\bar{s}(0)\gamma_{\alpha}D_{\beta}(\eta n)s(\lambda n)]\ ,\nn
J^1_{\alpha\beta}(0,\eta n,\lambda n)&=&\frac{1}{\sqrt{3}}[\bar{u}(0)\gamma_{\alpha}D_{\beta}(\eta n)u(\lambda n)+\bar{d}(0)\gamma_{\alpha}D_{\beta}(\eta n)d(\lambda n)+\bar{s}(0)\gamma_{\alpha}D_{\beta}(\eta n)s(\lambda n)]\ ,\nn
J^8_{\alpha\beta 5}(0,\eta n,\lambda n)&=&\frac{1}{\sqrt{6}}[\bar{u}(0)\gamma_5\gamma_{\alpha}D_{\beta}(\eta n)u(\lambda n)+\bar{d}(0)\gamma_5\gamma_{\alpha}D_{\beta}(\eta n)d(\lambda n)-2\bar{s}(0)\gamma_5\gamma_{\alpha}D_{\beta}(\eta n)s(\lambda n)]\ ,\nn
J^1_{\alpha\beta 5}(0,\eta n,\lambda n)&=&\frac{1}{\sqrt{3}}[\bar{u}(0)\gamma_5\gamma_{\alpha}D_{\beta}(\eta n)u(\lambda n)+\bar{d}(0)\gamma_5\gamma_{\alpha}D_{\beta}(\eta n)d(\lambda n)+\bar{s}(0)\gamma_5\gamma_{\alpha}D_{\beta}(\eta n)s(\lambda n)]\ .
\eee
Due to the factor $1-2x$ for $\tilde{G}_{P_i}$, $G_{P_i}$ become dominate. The normalizations of $\phi_P^i$ and $G_P^i$ are determined from the leptonic weak decay and the axial anomaly for $P_i$ meson, respectively. This is similar to the pion case \cite{twy1}.

\section{The Mixing Schemes}
We employ the $SU(3)_F$ octet and singlet states to describe the $\eta-\eta^\prime$ system. The $\eta$ and $\eta^\prime$ meson states can be described by means of the octet and singlet states $|\eta_8\rangle$ and $|\eta_1\rangle $ through the one mixing angle scheme
\bee\label{stmix}
|\eta\rangle &=&\cos\theta|\eta_8\rangle-\sin\theta|\eta_1\rangle\nn
|\eta^\prime\rangle &=&\sin\theta|\eta_8\rangle+\cos\theta|\eta_1\rangle
\eee
where the mixing angle $\theta$ controls the relative strength. With the mixing, the $\eta\gamma$ and $\eta^\prime\gamma$ form factors take expressions  
\bee\label{ffmix}
F_{\eta\gamma}(Q^2)&=&\cos\theta F_{\eta_8\gamma}(Q^2)-\sin\theta F_{\eta_1\gamma}(Q^2)\ ,\nn
F_{\eta^\prime\gamma}(Q^2)&=&\sin\theta F_{\eta_8\gamma}(Q^2)+\cos\theta F_{\eta_1\gamma}(Q^2)\ .
\eee 
To proceed, we also assume that the octet and singlet DAs take asymptotical form. Therefore, we have $\phi_{\eta_i}(x)=3f_{\eta_i} x(1-x)/\sqrt{2}$ and $G_{\eta_i}(x)=3\sqrt{2}\pi^2 f_{\eta_i}^3 x(1-x)$. 
The form factors $F_{\eta_i\gamma}(Q^2)$ (i=8,1) are simplified by substituting  $\phi_{\eta_i}(x)$ and $G_{\eta_i}(x)$
\bee\label{ffi}
F_{\eta_i\gamma}(Q^2)&=&4C_i \int dx \frac{1}{x(1-x)}[\frac{\phi_{\eta_i}(x)}{Q^2}-4\frac{G_{\eta_i}(x)}{Q^4}]\nn
&=&6\sqrt{2}C_i\frac{f_{\eta_i}}{Q^2}[1-\frac{8\pi^2f_{\eta_i}^2}{Q^2}] \ .
\eee 
Using Eq.~(\ref{ffi}) into Eq.~(\ref{ffmix}), we may derive the coefficients $a_P^{P_i}$ in Eq.~(\ref{nloff}). 

To compare the form factors with the data, we extrapolate the form factors to all orders 
\bee
F_{\eta_i\gamma}(Q^2)=\frac{6\sqrt{2}C_if_{\eta_i}}{Q^2+8\pi^2f_{\eta_i}^2}\ .
\eee
This formula gives a theoretical support to the approach using the interpolating formula for the $\eta\gamma$ and $\eta^\prime\gamma$ form factors \cite{Feldmann:1998yc}. 

The decay constants $f_{\eta_i}, i=8,1$ and the mixing angle $\theta$ will be determined by a least $\chi^2$ fit to the transition form factor data above $1$ GeV$^2$ and the two photon decay widths \cite{PDG2000}
\bee
\Gamma[\eta\to\gamma\gamma]=(0.46\pm 0.04)\;\;\makebox{keV}\;\;\ ,\;\;
\Gamma[\eta^\prime\to\gamma\gamma]=(4.28\pm 0.19)\;\;\makebox{keV}\ .
\eee
The decay rates have theoretical expressions   
\bee\label{dw}
\Gamma[\eta\to\gamma\gamma]&=&\frac{9\alpha^2}{32\pi^3}M_\eta^3\left[\frac{C_8 f_{\eta^\prime}^1-C_1 f_{\eta^\prime}^8}{f_{\eta^\prime}^1 f^8_\eta-f^8_{\eta^\prime}f_\eta^1}\right]^2\ ,\nn
\Gamma[\eta^\prime\to\gamma\gamma]&=&\frac{9\alpha^2}{32\pi^3}M_{\eta^\prime}^3\left[\frac{-C_8 f_{\eta}^1+C_1 f_{\eta}^8}{f_{\eta^\prime}^1 f^8_\eta-f^8_{\eta^\prime}f_\eta^1}\right]^2\ ,
\eee
where decay constants are defined as 
\bee
f_{\eta}^8&=&f_{\eta_1}\cos\theta\;\; ,\;\; f_{\eta}^1=-f_{\eta_1}\sin\theta\ ,\nn
f_{\eta^\prime}^8&=&f_{\eta_1}\sin\theta\;\; ,\;\; f_{\eta^\prime}^1=f_{\eta_1}\cos\theta\ .
\eee
The $\chi^2$ fit results are shown as Fit I in Figs.~5 and 6 and in Table.~I. It is seen that the Fit I is in good agreement with the data for the form factors. To test the fit parameters, we employ the ratio of the decay rates for $J/\psi$ into $\eta^\prime\gamma$ and $\eta\gamma$
\bee
R_{J/\psi}=\frac{\Gamma(J/\psi\to\eta^\prime\gamma)}{\Gamma(J/\psi\to\eta\gamma)}=5.0\pm 0.6.
\eee 
It is usually assumed that the radiative $J/\psi\to\eta(\eta^\prime)\gamma$ decays are dominated by nonperturbative gluon matrix elements $\langle 0|G\tilde{G}|\eta^\prime\rangle$ and $\langle 0|G\tilde{G}|\eta\rangle$ such that the ratio takes expression \cite{Kiselev:1993ms}
\bee
R_{J/\psi}=\left(\frac{M_{\eta^\prime}^2(f_{\eta^\prime}^8 + \sqrt{2} f_{\eta^\prime}^1 )}{M_{\eta}^2(f_{\eta}^8 + \sqrt{2} f_{\eta}^1 )}\right)^2\left(\frac{p_{\eta^\prime}}{p_{\eta}}\right)^3
\eee
with $p_P=M_{J/\psi}(1-M_P^2/M_{J/\psi}^2)/2$ being the three momentum of the $P$-meson. 
Our fit result is close to that one obtained from chiral perturbation theory ($\chi$PT), except the octet decay constant $f_{\eta_8}=f_\pi < 1.28 f_\pi$,
\bee\label{chipt1}
\chi\makebox{PT}: \theta\approx -20^\circ\sim-10^\circ \;\;\ ;\;\; f_{\eta_8}=1.28f_\pi
\;\;\ ;\;\; f_{\eta_1}\approx 1.1 f_\pi\ .
\eee
The octet decay constant $f_{\eta_8}$ is calculable by $\chi$PT up to one loop approximation
\bee
f_{\eta_8}\approx [1-\frac{M_K^2}{(4\pi f_\pi)^2}\ln\frac{M_K^2}{(4\pi f_\pi)^2}+\frac{M_\pi^2}{(4\pi f_\pi)^2}]f_\pi=1.28 f_\pi\ .
\eee
It is noted that the predict results for $\Gamma(\eta\to \gamma\gamma)$, $\Gamma(\eta^\prime\to \gamma\gamma)$ and $R_{J/\psi}$ shown in Table I are close to the experimental values within $1\sigma$ accuracy. 

Recently, it has been proposed \cite{Leutwyler:1998yr,Feldmann:1998vc,Feldmann:1998yc} that $|\eta\rangle $ and $|\eta^\prime\rangle$ can mix through a two mixing angle scheme as 
\bee\label{2stmix}
|\eta\rangle &=&\cos\theta_8 |\eta_8\rangle-\sin\theta_1|\eta_1\rangle\ ,\nn
|\eta^\prime\rangle &=&\sin\theta_8|\eta_8\rangle+\cos\theta_1|\eta_1\rangle\ ,
\eee
where $\theta_i$ denote the mixing angles. Using this mixing scheme, the $\eta\gamma$ and $\eta^\prime\gamma$ form factors can be expressed, correspondingly, as 
\bee\label{2ffmix}
F_{\eta\gamma}(Q^2)&=&\cos\theta_8 F_{\eta_8\gamma}(Q^2)-\sin\theta_1 F_{\eta_1\gamma}(Q^2)\ ,\nn
F_{\eta^\prime\gamma}(Q^2)&=&\sin\theta_8 F_{\eta_8\gamma}(Q^2)+\cos\theta_1 F_{\eta_1\gamma}(Q^2)\ .
\eee 
The form factors $F_{\eta_i\gamma}(Q^2)$ are the same as those in the one mixing angle scheme. We first change the values of $f_{\eta_i}$ and $\theta_i$ to fit the data. The $\chi^2$ fit result is shown as Fit II in Figs.~5 and 6 and Table.~I. From Fit I and Fit II in Table I, it is found that the two mixing angle scheme is better in accuracy than the one mixing angle scheme by 100$\%$. This is close to the investigations \cite{Leutwyler:1998yr,Feldmann:1998vc,Feldmann:1998yc}. 

A large value of $f_{\eta^\prime}^c=\cos\theta_1 f_{\eta_c}$ \cite{Halperin:1997as,Cheng:1997if}, which is responsible for the intrinsic charm content of the $\eta^\prime$ meson, has been proposed to resolve the large branching ratios Br$(B\to\eta^\prime K)$ and Br$(B\to X_s\gamma)$. We may explore this within our approach by adding intrinsic charm content into our formalism. The effects of the intrinsic charm content are similar to that of the singlet component. That means one can replace $f_{\eta_1}$ with $f_{\eta_c}=f_{\eta^\prime}^c/\cos\theta_1$, the decay constant for intrinsic charm for the corresponding singlet terms. That is the part of the form factors from the intrinsic charm taking expression
\bee
F_{\eta_c\gamma}(Q^2)=4e_c^2\int\frac{dx}{x(1-x)}\left[\frac{\phi_{\eta_c}(x)}{Q^2}-4\frac{G_{\eta_c}(x)}{Q^4}\right]
\eee
where $e_c=2/3$ and the related DAs are $\phi_{\eta_c}(x)=3f_{\eta_c} x(1-x)/\sqrt{2}$ and $G_{\eta_c}(x)=3\sqrt{2}\pi^2f_{\eta_c}^2x(1-x)$. The effects from the large value of the charm quark mass has been absorbed into the twist-4 DA $G_{\eta_c}(x)$\cite{twy1}. After including the contribution of the intrinsic charm, the $\eta\gamma$ and $\eta^\prime\gamma$ form factors then become 
\bee\label{2ffcmix}
F_{\eta\gamma}(Q^2)&=&\cos\theta_8 F_{\eta_8\gamma}(Q^2)-\sin\theta_1 (F_{\eta_1\gamma}(Q^2)+F_{\eta_c\gamma}(Q^2))\ ,\nn
F_{\eta^\prime\gamma}(Q^2)&=&\sin\theta_8 F_{\eta_8\gamma}(Q^2)+\cos\theta_1 (F_{\eta_1\gamma}(Q^2)+F_{\eta_c\gamma}(Q^2))\ .
\eee 
As the case of the octet and singlet form factors, the extrapolation of $F_{\eta_c\gamma}(Q^2))$ to all orders is implied. By observing Eq.~(\ref{2ffcmix}), form factor $F_{\eta^\prime\gamma}(Q^2)$ has a larger dependence of $f_{\eta_c}$ than $F_{\eta\gamma}(Q^2)$. We make a least $\chi^2$ fit to the form factor data to determine possible values of $f_{\eta_c}$ by keeping other parameters fixed. From Table II, one may see that including the intrinsic charm content can indeed improve the accuracy. This shows that our formalism is consistent in perturbation theory that the higher Fock state can be reasonably added in. As shown below that the allowed value for $f_{\eta_c}$ is less than $f_{\pi}$. In literature \cite{Halperin:1997as,Cheng:1997if,Feldmann:1998vc,Feldmann:1998yc}, $f_{\eta_c}$ is proposed in the range $-140$ MeV $\le f_{\eta_c} \le$ 15 MeV. To test this, we plot in Fig.~7 the $\chi^2$ distribution  for each set of mixing parameters list in Table II over a wide range of $-140$ MeV $\le f_{\eta_c} \le 140$ MeV. It is seen that the range of $f_{\eta_c}$: $-10$ MeV $\le f_{\eta_c} \le$ -4 MeV are allowed by the data.  Because the value of $\cos\theta_1$ is close to unity, $f_{\eta^\prime}^c$ is almost equal to $f_{\eta_c}$.

From the above analysis, one may see that combining the high energy data and the low energy experiment can result in constraints on the mixing parameters in a very efficient way. We can give a general analysis for the two mixing angle scheme. We shall investigate the $\chi^2$ distributions of the mixing parameters. The procedure of analysis is as followed. We firstly separate the data into two groups. The data for the form factors, the two photon decay rates and the ratio for $J/\psi$ radiative decays are denoted as set I while the latter two data are chosen as set II. We then determine the least $\chi^2$ values for sets I and II, respectively. The reason for separating the data into set I and set II is that neither set I nor set II can completely constraint the parameters. The parameters locate in $1\sigma$ accuracy of the set I still have large uncertainties and require further restrictions, which can be obtained by the data set II. To be more explicit, we plot the allowable regions for the mixing parameters within $1\sigma$ error with respect to those values associated with the minimal $\chi^2$ points. As shown in Figs.~8 and 9, both allowable regions for data set I and II are large while their intersections are rather restricted.  The reason for this fact is easily understood. Within the data set I, the experimental errors are shared for high and low energy data. Because the $\chi^2$ distribution can only measure the correlations between the mixing parameters, the hope for constraining each parameters in a independent way can not be obtained and only partial restrictions over the correlations of the parameters can be derived. This can be seen from Figs.~8, in which the $f_8$ parameter is not contrained in a reasonable way. To compensate this flaw, we note that the $\chi^2$ distribution of the data set II can intersect with that $\chi^2$ distribution of the data set I. The intersections between two $\chi^2$ distributions can give better constraints on parameters than the data set I or II. One should note that although the data set II is a subset of the data set I, the $\chi^2$ distribution of the data set II is not necessary to be a  subset of the $\chi^2$ distribution of the data set I, since the effects of correlations between the parameters are different for two data sets. From the overlapped regions of the data set I and II in $1\sigma$ error, we may extract from Figs.~8 and 9 the allowable regions for the mixing parameters. In Fig.~8, we plot the possible allowable ranges for $f_{\eta_8}$ and $f_{\eta_1}$. It can be observed from Fig.~8 that the overlapped region for $f_{\eta_8}$ and $f_{\eta_1}$ is quite stringent: $1.1\le f_{\eta_8}/f_\pi\le 1.56$ and $1.1\le f_{\eta_1}/f_\pi\le 1.22$. Figure 9 shows the allowable region for $\theta_1$ and $\theta_8$ from both data set I and set II. The overlapped region indicates that: $-12^\circ\le \theta_1 \le -6^\circ$ and $-23.6^\circ\le \theta_8\le -19^\circ$. 

Combining Figs.~8 and 9, we may derive the $Q^2\to\infty$ limits of the scaled $\eta\gamma$ and $\eta^\prime\gamma$ form factors 
\bee
Q^2F_{\eta\gamma}(Q^2)|_{Q^2\to\infty}=\frac{2}{\sqrt{3}}[ f_{\eta_8}\cos\theta_8-2\sqrt{2}f_{\eta_1}\sin\theta_1]=(189\pm 46)\makebox{MeV}\ , \nn
Q^2F_{\eta^\prime\gamma}(Q^2)|_{Q^2\to\infty}=\frac{2}{\sqrt{3}}[ f_{\eta_8}\sin\theta_8+2\sqrt{2}f_{\eta_1}\cos\theta_1]=(295\pm 35)\makebox{MeV}\ .
\eee
The error in the scaled $\eta\gamma$ form factor being larger than that of the scaled $\eta^\prime\gamma$ form factor is due to the fact that the errors in the data for $\eta\gamma$ and $\eta^\prime\gamma$ form factors are shared in our analysis. This is consistent with the concept of $\eta-\eta^\prime$ mixing. 

\section{Conclusions}
We have shown that the collinear expansion for $\gamma^*\eta(\eta^\prime)\to\gamma$ can be systematically performed in compatibility with PQCD factorization.
The $O(Q^{-4})$ power corrections for $F_{\eta\gamma}(Q^2)$ and $F_{\eta^\prime\gamma}(Q^2)$ have been evaluated. The magnitudes of NLO power corrections are determined. 

We have made a general analysis for the allowed values for the mixing parameters by combining the high and low energy data. Except of $f_{\eta_8}$, the other three parameters $f_{\eta_{1}}, \theta_{8(1)}$ can be constrained in a reasonable region. The large error for the fit $f_{\eta_8}$ is mainly from the experimental error and can be improved by future experiment with higher accuracy. At present, we would invoke the chiral perturbation theory calculation for $f_{\eta_8}$. 

We have also shown that the intrinsic charm content of $\eta^\prime$ meson has little contributions. Any sizable contribution from the intrinsic charm would lead to large $\chi^2$ value as shown in Fig.~7. Of course, there require further investigations for this point, if more accurate form factors data are available. 

So far, we have not considered the finite meson mass effects for the $\eta\gamma$ and $\eta^\prime\gamma$ form factors. At first glance, the meson mass effects can not be safely ignored. Because the values of $\eta$ and $\eta^\prime$ meson masses are large as $M_{\eta}=547$ MeV and $M_{\eta^\prime}=958$ MeV, power corrections from the mass effects of order $O(M_{P}^2/Q^2)$ and $O(M_P^2\Lambda^2/Q^4)$ are also important. The type of corrections $O(M_{P}^2/Q^2)$ are kinematics. This is similar to the Nachtmann's correction for deep inelastic scattering \cite{Nachtmann:1973mr} and can be negligible. The second type of corrections $O(M_P^2\Lambda^2/Q^4)$ are dynamics and can be argued that they are of twist-6, at least. This is because the associated spin projector is $\s{n}\gamma_5$ (cf. the leading spin projector $\s{p}\gamma_5$), which will introduce two additional $F_S$ propagators into the related hard function. As a result, the dynamical type meson mass corrections are of order $O(M_{P}^2\Lambda^4/Q^6)$. Of course, a complete analysis for these meson mass effects is important.


\noindent
{\bf Acknowledgments:}
This work was supported in part by the National
Science Council of R.O.C. under the Grant No. NSC89-2811-M-009-0024.

\newpage
\begin{center}
\begin{table}[h]
\begin{tabular}[b]{|c||c|c|c|c||c|c||c||c|}
\hline\hline
&$f_{\eta_8}/f_\pi$&$f_{\eta_1}/f_\pi$&$\theta_8$&$\theta_1$&$\Gamma(\eta\to 2\gamma)$[keV]&$\Gamma(\eta^\prime\to 2\gamma)$[keV]& $R_{J/\psi}$&$\chi^2/$dof\\
\hline\hline
Fit I& 0.99 & 1.08 & $-16.4^\circ$ &$-16.4^\circ$ &0.49 & 4.47 & 5.6 &64/31 \\
\hline
Fit II& 1.32 & 1.16 & $-22.3^\circ$ & $-9.1^\circ$ &0.50 & 4.34 & 4.4 & 32/31 \\
\hline \hline 
\end{tabular} 
\caption{The results of the $\chi^2$ fit to the $\eta\gamma$ and $\eta^\prime\gamma$ transition form factors and the two photon widths within the one and two mixing angle schemes. }
\end{table}
\end{center}

\begin{center}
\begin{table}[h]
\begin{tabular}[b]{|c||c|c|c|c|c||c|}
\hline\hline
&$f_{\eta_8}/f_\pi$&$f_{\eta_1}/f_\pi$&$f_{\eta_c}$[MeV]&$\theta_8$&$\theta_1$&$\chi^2/$dof\\
\hline\hline
Fit I & 0.99 & 1.08 & 0 & $-16.4^\circ$ & $-16.4^\circ$ &61/29 \\
\hline
Fit I & 0.99 & 1.08 & -8 & $-16.4^\circ$ & $-16.4^\circ$ &34/29 \\
\hline
Fit II & 1.32 & 1.16 & 0 & $-22.3^\circ$ & $-9.1^\circ$ &31/29 \\
\hline
Fit II & 1.32 & 1.16 & -5.6 & $-22.3^\circ$ & $-9.1^\circ$ &19/29 \\
\hline\hline 
\end{tabular} 
\caption{The results of the $\chi^2$ fit to the $\eta\gamma$ and $\eta^\prime\gamma$ transition form factors within the one and two mixing angle scheme with or without intrinsic charm content.}
\end{table}
\end{center}
\newpage
\begin{figure*}
\includegraphics{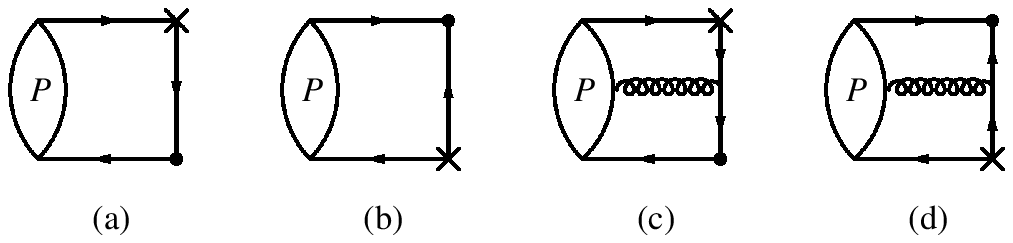}
\caption{\label{fig:fig1}The leading order diagrams for $\gamma^*\eta(\eta^\prime)\to\gamma$. The cross symbol means the vertex of the virtual photon.}

\includegraphics{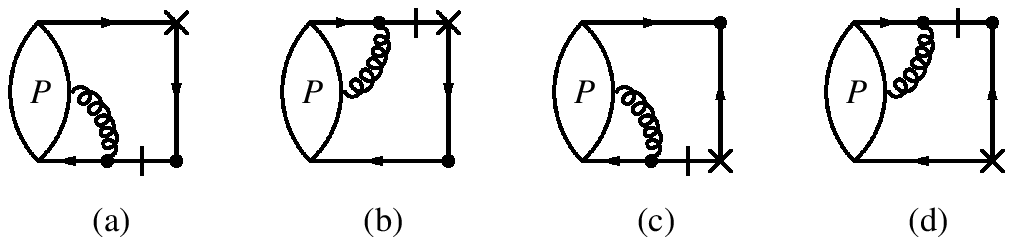}
\caption{\label{fig:fig2}The next-to-leading twist (NLT) diagrams for $\gamma^*\eta(\eta^\prime)\to\gamma$. The propagator with one bar means the special propagator.}
\end{figure*}
\begin{figure*}
\includegraphics{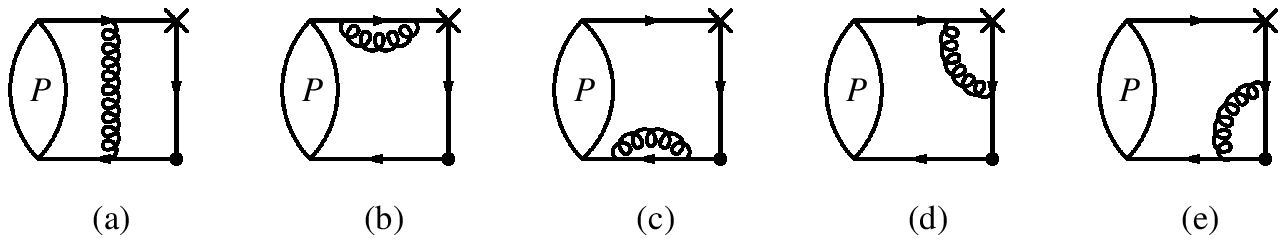}
\caption{\label{fig:fig3}The one loop diagrams for the LT amplitude of process $\gamma^*\eta(\eta^\prime)\to\gamma$.}

\includegraphics{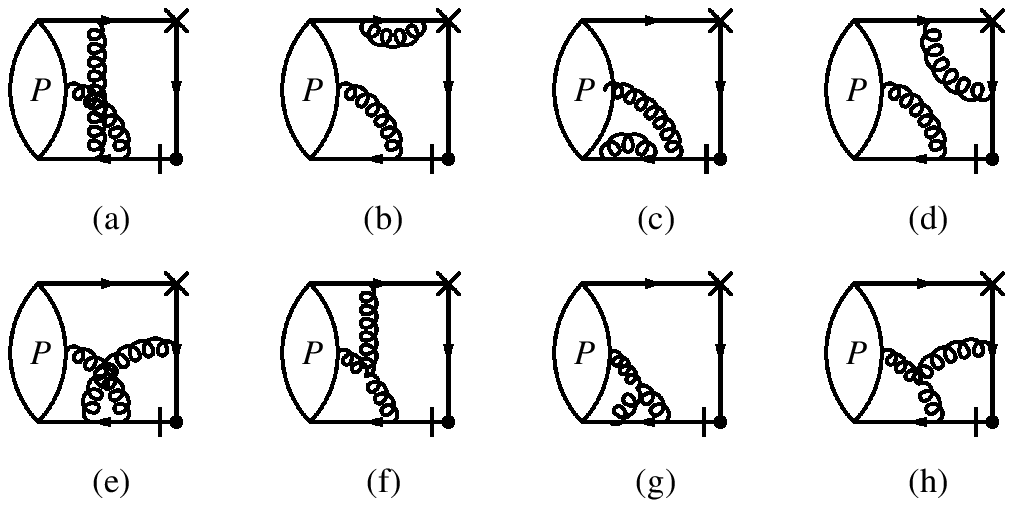}
\caption{\label{fig:fig4}The one loop diagrams for the NLT amplitude of process $\gamma^*\eta(\eta^\prime)\to\gamma$.}
\end{figure*}
\newpage
\begin{figure*}
\includegraphics{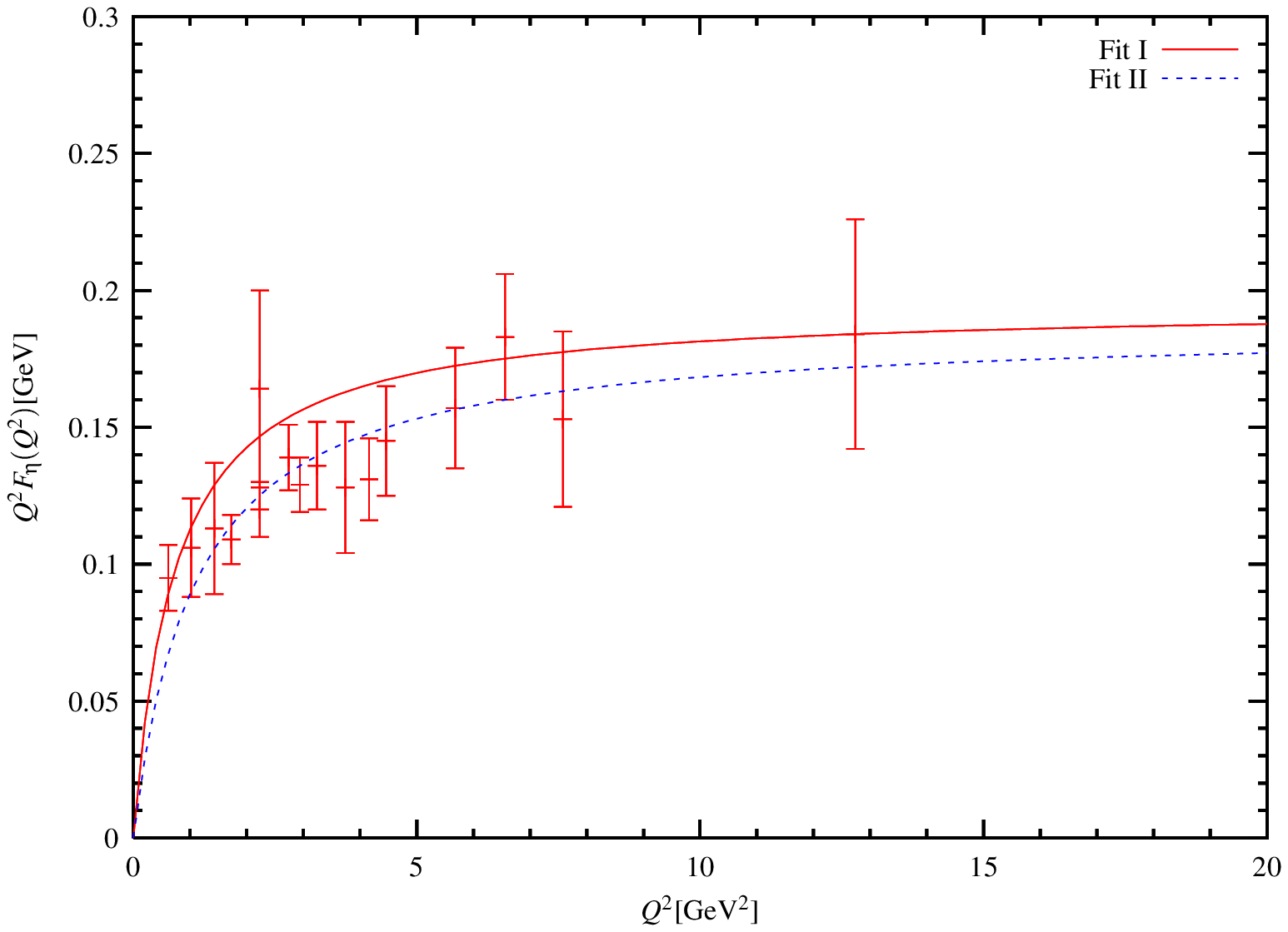}
\caption{\label{fig:fig5}The results of the $\chi^2$ fit to the $\eta\gamma$  transition form factor within the one mixing angle scheme (the solid line) and the two mixing angle scheme (the dashed line). The data point are taken from \cite{Gronberg:1998fj,Acciarri:1998yx,Behrend:1991sr,Aihara:1990nd,Berger:1984xk}.}
\end{figure*}
\newpage
\begin{figure*}
\includegraphics{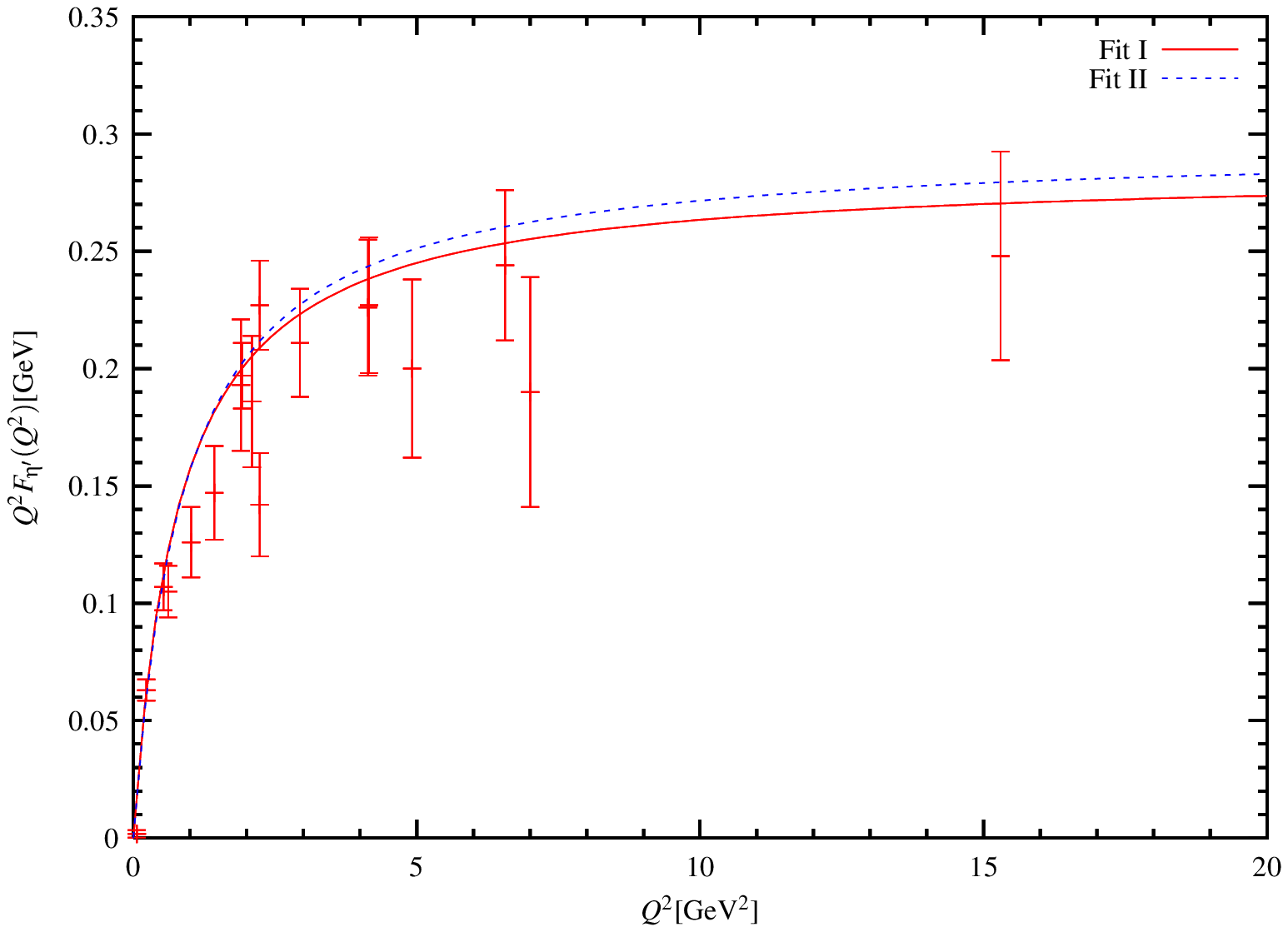}
\caption{\label{fig:fig6}The results of the $\chi^2$ fit to the $\eta^\prime\gamma$ transition form factor within the one mixing angle scheme (the solid line) and the two mixing angle scheme (the dashed line). The data point are taken from \cite{Gronberg:1998fj,Acciarri:1998yx,Behrend:1991sr,Aihara:1990nd,Berger:1984xk}.}
\end{figure*}
\newpage
\begin{figure*}
\includegraphics{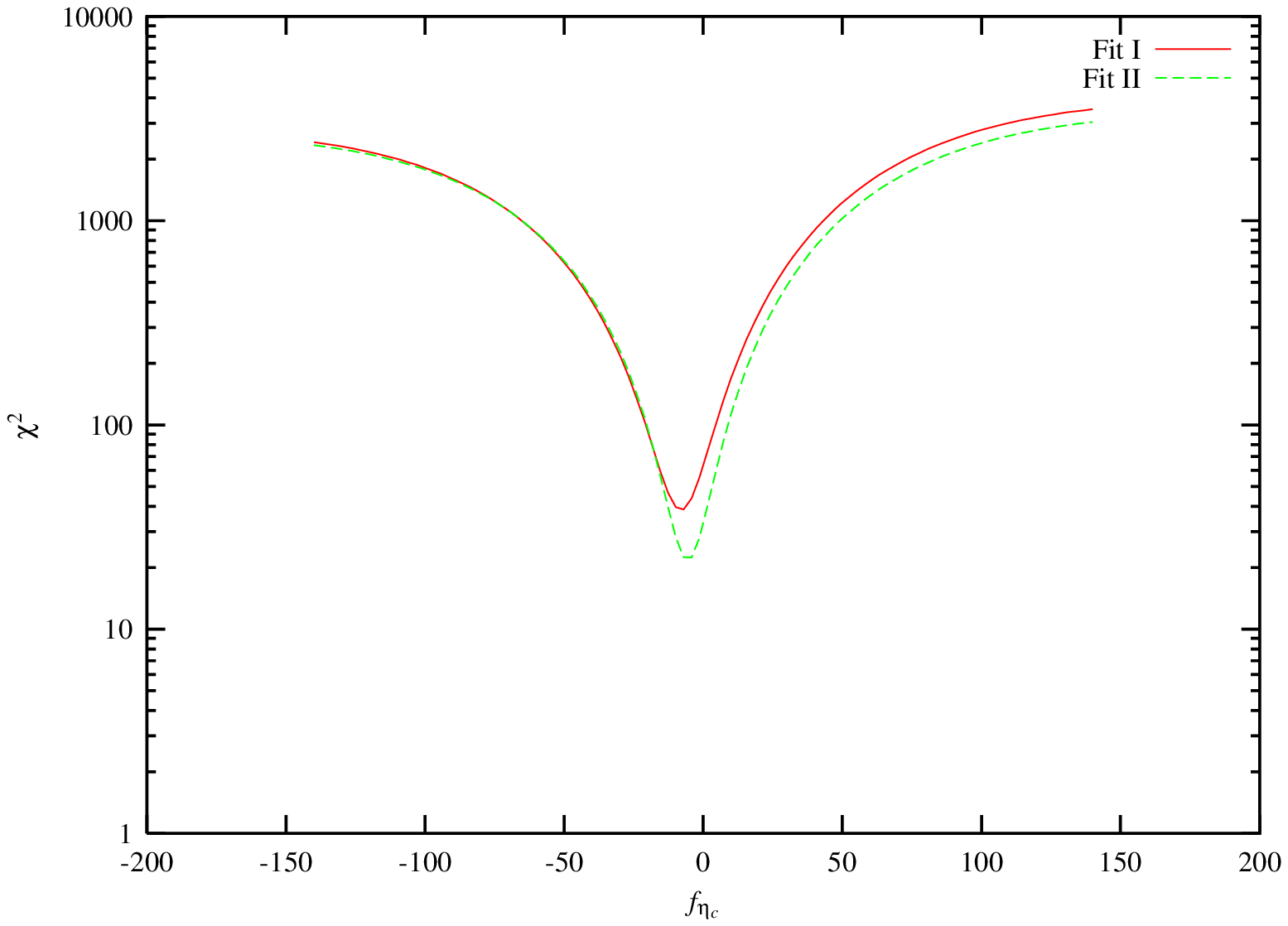}
\caption{\label{fig:fig7}The plot of the $\chi^2$ distribution vs. $f_{\eta_c}$. The $\chi^2$ values are evaluated for the data for forms factors by employing the sets of parameters list in Table I.}
\end{figure*}
\begin{figure*}
\caption{\label{fig:fig8}The plot of $f_{\eta_1}$ versus $f_{\eta_8}$ from the $\chi^2$ analysis for the data set I and II. The regions denoted as I and II represent the allowable values for $f_{\eta_1}$ and $f_{\eta_8}$ within $1\sigma$ error for corresponding data set I and II.
}
\end{figure*}
\newpage
\begin{figure*}
\caption{\label{fig:fig9}The plot of $\theta_1$ versus ${\theta_8}$ from the $\chi^2$ analysis for the data set I and II. The regions denoted as I and II represent the allowable values for $\theta_1$ and ${\theta_8}$ within $1\sigma$ error for corresponding data set I and II.
}
\end{figure*}

\end{document}